\documentclass[prc,twocolumn,superscriptaddress,nofootinbib,showkeys]{revtex4}

\usepackage{amsmath,amssymb}

\usepackage{mathrsfs}
\usepackage{color}
\usepackage{graphicx}

\usepackage{epsfig,psfrag}
\usepackage{array}

\begin{document}
\title{\boldmath N3LO $NN$ interaction adjusted to light nuclei in {\slshape ab exitu} approach}
\author{A. M. Shirokov}
\affiliation{Skobeltsyn Institute of Nuclear Physics, Moscow State University,
Moscow, 119992, Russia}
\affiliation{Department of Physics and Astronomy,
Iowa State University, Ames, IA 50011-3160, USA}
\affiliation{Pacific National University, 136 Tikhookeanskaya st., Khabarovsk 680035, Russia}
\author{I. J. Shin}
\affiliation{{Rare Isotope Science Project, Institute for Basic Science, Daejeon 34037, Korea}}
\author{{Y. Kim}}
\affiliation{{Rare Isotope Science Project, Institute for Basic Science, Daejeon 34037, Korea}}
\author{M. Sosonkina}
\affiliation{Department of Modeling, Simulation and Visualization Engineering, Old Dominion University, Norfolk, VA 23529, USA}
\author{P. Maris}
\affiliation{Department of Physics and Astronomy,
Iowa State University, Ames, IA 50011-3160, USA}
\author{J. P. Vary}
\affiliation{Department of Physics and Astronomy,
Iowa State University, Ames, IA 50011-3160, USA}

\begin{abstract}
 We use phase-equivalent transformations to adjust off-shell properties of similarity renormalization group evolved
chiral effective field theory 
$NN$ interaction (Idaho N3LO)  to fit selected binding energies and spectra of light nuclei in an {\em ab exitu} approach. 
We then test the transformed interaction on a set of additional observables in light nuclei to verify that it provides reasonable descriptions 
of these observables with an apparent reduced need for three- and many-nucleon interactions.
\end{abstract}


\maketitle

An {\em ab initio} description of nuclear structure and reactions is one of the mainstreams of modern nuclear theory~\cite{LiedOrl}. 
It is based on a rapid development of
supercomputer facilities and recent advances in the utilization of high-performance computing
systems~\cite{Bogner:2013pxa}. Modern  {\em ab initio} nuclear theory  has opened a wide range of nuclear phenomena 
that can be evaluated to high precision using realistic nucleon-nucleon ($NN$) and three-nucleon ($NNN$) interactions. In particular,  
{\em ab initio} approaches,  such as
the No-Core Shell Model (NCSM)~\cite{NCSM-rev}, the Green's Function Monte Carlo (GFMC)~\cite{GFMC} and 
the Coupled-Cluster Theory~\cite{CC}, are able to reproduce properties of a large number of atomic nuclei with mass up to  
$A=16$ and 
selected heavier nuclear systems around closed shells. Very important progress has been achieved in the {\em ab initio} 
description of
reactions with light nuclei, in particular,  by combining the NCSM with the Resonating Group Method~\cite{NCSM-RGM-rev}.

The  {\em ab initio} theory requires a high-quality realistic inter-nucleon 
interaction providing an accurate description of $NN$
scattering data and predictions for binding energies, spectra and other observables in light nuclei. A number of 
meson-exchange potentials sometimes supplemented with phenomenological terms to achieve high accuracy in fitting $NN$
 data, e.\:g., CD-Bonn~\cite{CD-Bonn}, Nijmegen~\cite{Nijm}, Argonne~\cite{Argonne}, have been developed that should be used 
together with modern $NNN$ forces such as Urbana~\cite{Urbana, Urbana3}, Illinois~\cite{Illinois},  
Tucson--Melbourne~\cite{TM,TM2,TM3} to reproduce properties of 
many-body nuclear systems. 
A very important step in the theory of inter-nucleon interactions in nuclei is  the emergence of realistic $NN$ 
and $NNN$ interactions  tied to QCD via chiral effective field theory ($\chi$EFT)~\cite{EFT0, EFT1, N3LO, N4LO,NNN-EFT}.

Three-nucleon forces require a significant increase of computational resources in order to diagonalize a many-body 
Hamiltonian matrix since the $NNN$ interaction increases the number of non-zero matrix elements approximately by a factor of~30 in the case of $p$-shell nuclei~\cite{F30, MAris12C}.
As a result, one needs to restrict the basis space in many-body calculations when $NNN$ forces are involved which makes the predictions less precise. {\em Ab initio} 
many-body studies benefit from the use of recently developed purely two-nucleon interactions such as INOY 
(Inside Nonlocal Outside Yukawa)~\cite{INOY1,INOY2} and JISP ($J$-matrix Inverse Scattering 
Potential)~\cite{ISTP,JISP6, JISP6G, JISP16} types which are fitted not only to the $NN$ data but also to binding energies 
of~$A=3$ and heavier nuclei. 
At the fundamental level, these $NN$ interactions are supported by the work of Polyzou and Gl\"ockle~\cite{PolGl}
who demonstrated that a 
given $NN$ interaction is equivalent at the~$A=3$ level to 
some other $NN$ interaction augmented by $NNN$ interactions,
where 
the two $NN$ interactions are related  through a phase-equivalent transformation (PET). It seems reasonable then to exploit this freedom and 
strive to minimize the need for the explicit introduction of three- and higher-body forces. Endeavors along these lines have resulted in the design of INOY and JISP 
inter-nucleon interaction models. 

Conventional realistic meson-exchange $NN$ interactions~\cite{CD-Bonn,Nijm, Argonne} and $NN$ interactions
obtained 
via $\chi$EFT~\cite{N3LO} present convergence challenges in 
many-body calculations. A modern tool to soften the $NN$ interaction and hence to improve the convergence, is the
Similarity Renormalization Group (SRG) technique~\cite{Glazek:1993rc,Wegner:1994}. The SRG softening guarantees a monotonic
convergence of many-body calculations as 
a function of increasing basis space size and makes it possible to extrapolate the results to the infinite basis space
thus improving essentially an accuracy of theoretical predictions. We note that the SRG  softening of 
$NN$ interaction induces $NNN$ and, generally, four-nucleon ($4N$), five-nucleon, etc., forces.

We develop here an $NN$ interaction based on $\chi$EFT able to describe light nuclei without explicit use of $NNN$ forces and 
with good
convergence of many-body {\em ab initio} calculations. This interaction which we hereafter refer to as Daejeon16  $NN$ interaction should
be useful for 
a wide range of applications in nuclear structure and nuclear reactions.
We start from the Idaho N3LO $\chi$EFT $NN$ force~\cite{N3LO} SRG-evolved with
the flow parameter~$\lambda=1.5$~fm$^{-1}$ and apply to it various PETs with continuos parameters searching for an optimal set of
PET parameters providing a good description of light nuclei.
In our approach, we assume that our selected PETs 
are generating $NNN$ forces which cancel 
the combined effect of the `intrinsic' $NNN$ interaction and 
the $NNN$ force induced by the SRG transformation. 
Insofar as the PETs also provide a good fit to nuclei with $A = 4$, and beyond, we interpret that success as an indication that effects of neglected $4N$ forces, and beyond, are also minimized.

The technique used to construct the Daejeon16 interaction has much in common with the one utilized in constructing the 
JISP6~\cite{JISP6, JISP6G} and JISP16~\cite{JISP16} $NN$ interactions. In particular, we use the PETs of the same type~--- mixing lowest
components of the interaction matrix in the oscillator basis which were suggested in~\cite{PHT,LurAnnPh}. A minor difference is that these
PETs are utilized in the oscillator basis with the frequency~$\hbar\Omega=25$~MeV while~$\hbar\Omega=40$~MeV was used in the
JISP6 and JISP16 case. More important differences are the use of the SRG-evolved Idaho N3LO interaction instead of the ISTP interaction
of Ref.~\cite{ISTP} for PETs and a more accurate fitting to nuclear energies due to the use of the extrapolation technique of Ref.~\cite{ExtrapAB}
instead of a combination of results obtained with OLS-transformed and `bare' interaction in Refs.~\cite{JISP6,JISP6G,JISP16}.

We note here that the JISP16 $NN$ interaction appeared to be very successful in describing light nuclei (see a summary of the JISP16 results 
for $p$-shell nuclei in
Refs.~\cite{NN-NNNbook,IJMPE-2013}). In particular, the accuracies of $^{14}$F binding energy and spectrum  predictions ~\cite{14F} based on this interaction
were later confirmed by the first experimental study of this nucleus in Ref.~\cite{14Fexp}. However, the fit of the JISP16 interaction to light
nuclei was performed in 2006 with supercomputers of that era 
and hence within 
bases that 
are small by today's standards.  In addition, those calculations were performed  
without the use
of the extrapolation technique to the infinite model space which was introduced later. As a result, the JISP16 interaction was found to be
less accurate in the description of nuclei with mass $A>12$ and of some exotic light nuclei far away from $N=Z$ 
(see, e.~g.,
Refs.~\cite{NN-NNNbook,IJMPE-2013}).  We note also that JISP16 is a completely phenomenological $NN$ interaction whose design starts from
the inverse scattering fit to the $NN$ data~\cite{ISTP} without any underlying physics model. The Daejeon16 $NN$ interaction
is free from these drawbacks. Its fit to the many-body nuclear data is more accurate and it is obtained from the N3LO interaction of
Ref.~\cite{N3LO} by means of a well-defined SRG transformation and PETs. As a result, one can obtain the  effective operators,  e.~g., electroweak
 operators, that should be used in {\em ab initio} studies of many-body nuclear systems with Daejeon16 by applying the same
SRG  transformation and PETs to the `bare' operators consistent with the $\chi$EFT theory. 
We note that such SRG and PET transformations of the two-body chiral EFT electroweak operators can be included straightforwardly in future applications. It is also worth 
noting here that the 
SRG  transformation and PETs do not affect the description of $NN$ data and deuteron binding energy provided by the Idaho
N3LO $NN$ interaction.

We admit that, although it may be possible to weaken three-, four- and many-body interactions by performing PETs with fits to selected observables, 
one cannot in general eliminate them completely.  These interactions have a natural size in the context of chiral EFT suggesting that reduction below 
that size amounts to fine tuning which could succeed on a limited scale as we demonstrate here.  However, one anticipates that other observables, 
such as properties of heavier nuclei, may or may not be improved relative to experiment but further effort is needed to test such behaviors.

As noted above, we start from the  Idaho N3LO $\chi$EFT $NN$ interaction~\cite{N3LO}, SRG-evolve it with
the flow parameter~$\lambda=1.5$~fm$^{-1}$ and apply to it PETs of the type utilized in Ref.~\cite{ISTP,JISP6,JISP6G,JISP16} using
the oscillator basis with~$\hbar\Omega=25$~MeV. The PETs are mixing the lowest oscillator components of the wave function in each 
$NN$ partial wave; in case of coupled  $^{3}sd_{1}$ and $^{3}pf_{2}$ waves we mix by PETs the lowest $s$ and $p$ components, respectively.
The Daejeon16 $NN$ interaction is designed to be charge- and isospin-independent,
hence the $pn$ component 
after the PET
is used to obtain the Daejeon16 interaction in all $NN$ partial waves in $nn$, 
$pn$ and $pp$ channels; in the latter case it should be supplemented by the Coulomb interaction. 

The set of PET parameters in each partial wave is obtained by the fit to binding energies of $^{3}$H, $^{4}$He, $^{6}$Li, $^{8}$He, $^{10}$B, $^{12}$C
and $^{16}$O nuclei and to excitation energies of a few narrow excited states: 
the two lowest excited states with $(J^{\pi},T)=(3^{+},0)$ and $(0^{+},1)$ in $^{6}$Li and the first excited states $(1^{+},0)$ in $^{10}$B 
 and $(2^{+},0)$ in $^{12}$C. We minimize
the root-mean-square (rms) deviation of weighted differences of the calculated energies from target values using the POUNDerS derivative-free algorithm~\cite{pounders} as
implemented in~\cite{tao,PETSc}. The many-body calculations are performed within the NCSM using
the code 
MFDn~\cite{Maris_2010_2,Aktulga_2012,Concurrency2014}. To save computational resources, the minimization is performed using NCSM
calculations with relatively small basis spaces; the target values in the fit are the energies in respective nuclei in these small basis spaces
which are 
expected to result in correct experimental values after performing the extrapolations of Ref.~\cite{ExtrapAB} to the infinite model space.
The modification of the $NN$ interaction by PETs changes the convergence rate of NCSM calculations in each nucleus individually. 
Therefore after the initial fit we recalculate all targeted nuclei with the obtained interaction in a set of larger basis spaces, adjust the target energy 
values and perform a
new fit; recalculations in larger basis spaces with the new version of the $NN$ interaction result in a further adjustment of the target values
and in a new fit, etc.

\begin{table}
\vspace{-2ex}
\caption{PET angles (in degrees) defining the Daejeon16 $NN$ interaction in various $NN$ partial waves.}
\label{PETtable}
\begin{tabular}{c|ccccccc}\hline
Wave & $^{1}s_{0}$ &$^{3}sd_{1}$ & $^{1}p_{1}$ &$^{3}p_{0}$ &$^{3}p_{1}$ &$^{3}pf_{2}$ &$^{3}d_{2}$\\
Angle & $-2.997$ & 4.461 &5.507 &1.785 &4.299 &$-2.031$ &7.833\\
\hline
\end{tabular}
\end{table}

The  set of PET angles resulting from this multi-step fit and defining the Daejeon16 $NN$ interaction is presented in 
Table~\ref{PETtable}, the definition of these PET angles is given in Refs.~\cite{ISTP,JISP6, JISP6G, JISP16}\footnote{We note that we mix here
by PETs the two lowest $s$ components in the coupled  $^{3}sd_{1}$ waves while PETs utilized in Refs.~\cite{ISTP,JISP6, JISP6G, JISP16}
mix the lowest $s$ with the lowest $d$ components.}. 
The Daejeon16 interaction
is defined in all $NN$ partial waves with total angular momentum~$J\leq 6$;  the interaction in all partial waves not listed in Table~\ref{PETtable}
is 
the SRG-evolved Idaho N3LO interaction without a PET.  
For practical use, we refer to
a FORTRAN code generating the Daejeon16 $NN$ interaction matrix elements in the oscillator basis 
with~$\hbar\Omega=25$~MeV~\cite{Daejeon16Fort}.

\begin{figure}
\includegraphics[width= \columnwidth]{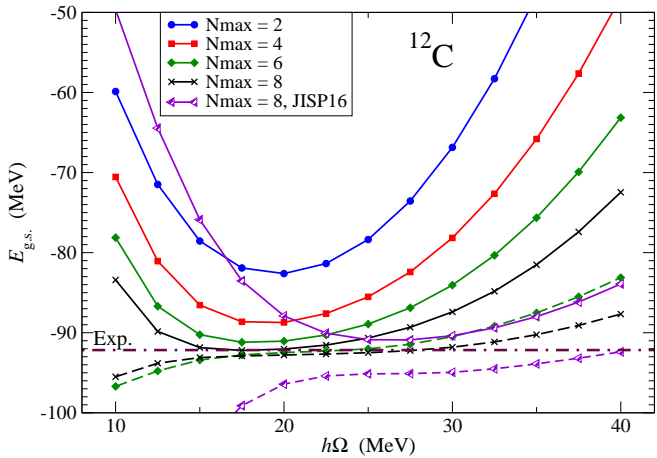}
\caption{$^{12}$C ground state energy in NCSM calculations obtained with Daejeon16 $NN$ interaction
with $N_{\max}$ values ranging from 2 to 8 as a function 
of~$\hbar\Omega$ (solid lines) and Extrapolation B results from basis spaces  up to respective $N_{\max}$ value (dashed lines). 
The~$N_{\max}=8$ results obtained with JISP16 $NN$ interaction are given for comparison. The horizontal dash-dotted line shows
the experimental $^{12}$C ground state energy~\cite{cdfe}.}
\label{12Cgs}
\end{figure}

{\em Ab initio} NCSM calculations with the Daejeon16 $NN$ interaction demonstrate a fast convergence as is illustrated by Fig.~\ref{12Cgs} where
we present the results obtained in NCSM basis spaces with excitation quanta~$N_{\max}$ ranging from~2 to~8 as functions of~$\hbar\Omega$. 
We show in Fig.~\ref{12Cgs} also the results of extrapolation to infinite NCSM model space for each~$\hbar\Omega$ (Extrapolations~B
of Ref.~\cite{ExtrapAB}) derived from the NCSM results from 3 successive basis spaces up to~$N_{\max}=6$ and~8. The NCSM results
are seen to 
converge as~$N_{\max}$ increases 
around the minimum of the~$\hbar\Omega$ dependence. This minimum 
for~$N_{\max}=6$ and~8 is very close to the extrapolated values and  the~$\hbar\Omega$ dependence is weak around the minimum. These
convergence patterns are due to the small value of the SRG flow parameter~$\lambda=1.5$~fm$^{-1}$ and are 
consistent with the study of convergence for various~$\lambda$ values of 
Ref.~\cite{BFMPSV2008}. For comparison, we show in Fig.~\ref{12Cgs} 
the~$N_{\max}=8$ results and respective extrapolations obtained with JISP16 interaction. JISP16 was designed to be a very soft interaction 
providing a fast convergence of {\em ab initio} studies. Nevertheless, it is seen that the extrapolated values are much farther from 
the NCSM JISP16
results than in the case of Daejeon16, hence the Daejeon16 $NN$ interaction provides a much better convergence than JISP16.

\begin{table}[!]
\vspace{-2ex}
\caption{Binding energies (in MeV) of nuclei obtained with Daejeon16 $NN$ interaction using Extrapolation B of Ref.~\cite{ExtrapAB}
with estimated uncertainty of the extrapolation (in parentheses),
the optimal $\hbar\Omega$  (in MeV) for the extrapolation and
the largest  $N_{\max}$ value used in NCSM calculations. The JISP16 results are given for comparison. 
The experimental data are taken from Ref.~\cite{cdfe}.}
\label{bind16}
\hspace*{-1ex}\begin{tabular}{cccccccc}\hline
& &\multicolumn{3}{c}{Daejeon16}&\multicolumn{3}{c}{JISP16}\\
\!\!\raisebox{1.2ex}[0pt][0pt]{Nucleus}\! &\raisebox{1.2ex}[0pt][0pt]{Nature}\! &Theory & {$\hbar\Omega$}  & { \!$N_{\max}$\!}   &Theory & {$\hbar\Omega$}  & { \!\!$N_{\max}$\!\!}   \\ \hline
$^{3\vphantom{\int}}$H & 8.482  & $8.442(^{+0.003}_{-0.000})$\strut &12.5  &16 &8.370(3) & 15 & 20\\[1pt]
$^3$He &7.718  &$7.744(^{+0.005}_{-0.000})$\strut &12.5  &16 &  7.667(5) & 17.5 & 20\\
$^4$He &28.296 &28.372(0) & 17.5 &16 & 28.299(0) & 22.5 &18\\
$^6$He &29.269  &29.39(3) &12.5  & 14 &28.80(5) &17.5 & 16\\
$^8$He &31.409  &31.28(1)&12.5  & 14 & 29.9(2) & 20 & 14\\
$^6$Li & 31.995 & 31.98(2)&12.5  & 14 & 31.48(3) & 20 & 16\\
$^{10}$B &64.751 &64.79(3) &17.5 &10  &63.9(1) & 22.5 &10 \\
$^{12}$C &92.162 &92.9(1) 
                                                     &17.5 &8 & 94.8(3) & 27.5 &10\\
$^{16}$O &127.619 &131.4(7) &17.5 &8 & 145(8) & 35 & 8\\ \hline
\end{tabular}
\end{table}

We present in Table~\ref{bind16} the extrapolated results of NCSM calculations of binding energies
of several $s$- and $p$-shell nuclei. 
Daejeon16 is
seen from Table~\ref{bind16} to provide an accurate description of these binding energies. In particular, Daejeon16 describes the
bindings generally better than the JISP16 $NN$ interaction whose  results are also shown in Table~\ref{bind16} for comparison. The main
drawbacks of the JISP16 interaction~--- overbinding of nuclei at the end of $p$ shell such as $^{16}$O and too strong decrease of binding
energies as~$|N-Z|$ increases, e.~g., underbinding of $^{6}$He and $^{8}$He~--- are much less pronounced in the case of
Daejeon16. As a manifestation of the fast convergence of Daejeon16 calculations, the same precision of binding energy 
extrapolations is achieved in smaller basis spaces as compared with JISP16.

\begin{table}[t!]
\vspace{-2ex}
\caption{Excitation energies (in MeV) of some nuclei obtained with Daejeon16 $NN$ interaction using Extrapolation~B of Ref.~\cite{ExtrapAB} 
with estimation of
uncertainty (in parentheses) for the absolute energy of the respective state, the optimal $\hbar\Omega$  
(in MeV) for the extrapolation of the excited state and
the largest  $N_{\max}$ value used in NCSM calculations. The JISP16 results are given for comparison.  
The experimental data are taken from Ref.~\cite{cdfe}.}
\label{excit16}
\begin{tabular}{cccccccc}\hline
Nucleus,& &\multicolumn{3}{c}{Daejeon16}&\multicolumn{3}{c}{JISP16}\\
level &\raisebox{1.2ex}[0pt][0pt]{Nature} &Theory & {$\hbar\Omega$} & { $N_{\max}$} &Theory & {$\hbar\Omega$} & { $N_{\max}$}  \\ \hline
$^{6\vphantom{\int}}$He \\
$(0^{+},1)$ & 0 & 0 & & & 0\\
$(2^{+},1)$ & $1.797$
                                                                                 & 1.91(5) &12.5 & 14 & 2.3(1) & 17.5 & 16 \\[3pt]
$^6$Li \\
$(1^{+}_{1},0)$ & 0 & 0 & & & 0\\\
$(3^{+},0)$ & $2.186$
                                                                                 &1.91(1) & 12.5 &14 & 2.55(7) & 20 & 16\\
$(0^{+},1)$ & $3.563$
                                                                                 & 3.50(4) & 12.5 &14 & 3.65(6) &17.5 & 16 \\
$(2^{+},0)$ & $4.312$
                                                                                 & 4.4(3) & 12.5 &14 & 4.5(2) & 20 &16 \\
$(2^{+},1)$ & $5.366$
                                                                                  & 5.36(7)& 12.5 &14 & 5.9(1) & 17.5 & 16 \\
$(1^{+}_{2},0)$ & $5.65$
                                                                                 & 5.0(4)& 12.5 &14  & 5.4(2) & 17.5 & 16 \\[3pt]
$^{10}$B \\
$(3^{+}_{1},0)$ & 0 & 0& & & 0\\
$(1^{+}_{1},0)$ &0.718
                                                                                  & 0.5(1) & 15 &10 & 0.9(2.4)  & 22.5 & 10\\
$(0^{+},1)$ & 1.740
                                                                                 & 1.74(7) & 17.5 & 10 & 1.8(1.4) &25 & 8\\
$(1^{+}_{2},0)$ & 2.154
                                                                               & 2.8(2)   & 17.5 & 10 & 4.1(1.7) & 30 &10\\
$(2^{+},0)$ & 3.587
                                                                                 & 4.3(2) & 15 &10 & 3.8(2) & 27.5 & 10 \\
$(3^{+}_{2},0)$ & $4.774$
                                                                                 & 5.1(7) & 17.5 & 10 & 5.6(3) & 22.5 & 10 \\
$(2^{+},1)$ & $5.164$
                                                                                 & 5.49(9)  & 17.5 & 10 & 4.6(3) & 22.5 & 10\\[3pt]
$^{12}$C \\
$(0^{+},0)$ & 0 & 0& & & 0\\
$(2^{+},0)$ & 4.439 & 4.57(15) &17.5 &8  & 3.9(4) & 27.5 & 10\\ \hline
\end{tabular}
\end{table}

Spectra of $^{6}$Li and $^{10}$B together with excitation energies of the first excited states in $^{6}$He and $^{12}$C are shown in 
Table~\ref{excit16}. Note, only two lowest narrow excited states in $^{6}$Li and the first excited states in $^{10}$B and $^{12}$C were involved in the fit.
We calculate the uncertainties of excitation energies as uncertainties of extrapolations of absolute energies of respective levels. The
uncertainties of the absolute energies include 
 the uncertainty of the overall  binding energy 
and that uncertainty (listed in Table III) is conservative when quoted as the uncertainty for excitation energies due to cancellations of the systematic error contributions. Note that 
the  precision of excitation energies obtained with Daejeon16 is generally 
 better than the  precision
of the excitation energies obtained with JISP16 reflecting the faster convergence of {\em ab initio} calculations with Daejeon16.


The spectra of light nuclei shown in Table~\ref{excit16} are well reproduced by Daejeon16. The ordering of levels is correct with an exception
of a wide ($ \Gamma=1.5$~MeV) $(1^{+}_{2},0)$ state in $^{6}$Li for which ordering in the spectrum is uncertain due to large 
 error bars of its extrapolated energy overlapping the neighboring narrow $(2^{+},1)$ state.
We note that it is widely accepted  that the spin of the $^{10}$B ground state cannot be reproduced without an explicit use of
$NNN$ interactions. Our calculations with Daejeon16 
 demonstrate that  the $^{10}$B ground state spin can be obtained
using only two-nucleon interaction which, however, mimics the effects of $NNN$ forces by modification of its off-shell properties by means
of PETs. 
The correct spin of the $^{10}$B ground state may also  be reproduced by the JISP16 $NN$ interaction, however the
uncertainties of extrapolations of JISP16 results (see Table~\ref{excit16}) prevent a definitive  conclusion about the
ordering of the lowest $(3^{+}_{1},0)$ and $(1^{+}_{1},0)$ states in $^{10}$B.
%

\begin{figure}
\includegraphics[width= \columnwidth]{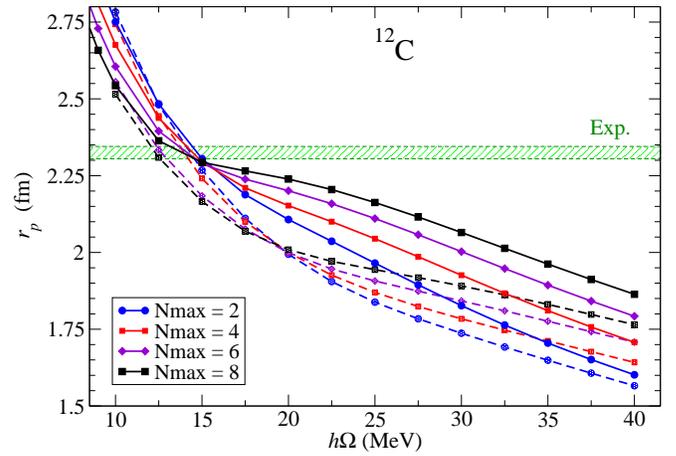}
\caption{$^{12}$C point-proton radius   as a function
of~$\hbar\Omega$ obtained with Daejeon16 (solid lines) and JISP16 (dashed lines) $NN$ interactions in NCSM calculations with 
various $N_{\max}$ values. The shaded area shows the experimental value with its uncertainties~\cite{AjS90}.}
\label{12Cr}
\end{figure}

The JISP16 interaction typically underestimates rms radii of nuclei.
As is seen in Fig.~\ref{12Cgs}, the~$\hbar\Omega$-dependence of NCSM eigenenergies obtained with Daejeon16 interaction
has a minimum at a much smaller~$\hbar\Omega$ value than in the case of the JISP16 $NN$ interaction. This is true not only for $^{12}$C
but also for other nuclei as is illustrated by Tables~\ref{bind16} and~\ref{excit16} where we present optimal~$\hbar\Omega$  values
for Extrapolation~B
which are close to the minima of respective $\hbar\Omega$-dependences. This feature suggests that the rms nuclear radii 
obtained with Daejeon16 will be closer to experiment since smaller~$\hbar\Omega$ values correspond to larger rms radii of
basis oscillator functions. 
The rms radii obtained in NCSM calculations are~$N_{\max}$ 
and~$\hbar\Omega$-dependent (see Fig.~\ref{12Cr} where we present the $^{12}$C point-proton rms radii~$r_{p}$ obtained with Daejeon16 interaction
in comparison with those from JISP16).
The~$\hbar\Omega$ dependencies of  rms radii obtained with a given interaction with different~$N_{\max}$ values tend to cross each other 
approximately at the same point. 
We use these crossing points 
as  rough estimates of the converged radius as has been suggested in Ref.~\cite{CMV2014}.
We note here that the extrapolation technique for nuclear rms radii was recently suggested~\cite{rp-extrap}. However, this technique was not
tested for nuclei with masses~$A>2$ and requires results from very large~$\hbar\Omega$ values ($\hbar\Omega> 49$~MeV were utilized in
Ref.~\cite{rp-extrap}) which were not used in our study. We note also that a detailed study of Ref.~\cite{Negoitathesis}
of the $^{12}$C point-proton rms radius with JISP16
interaction using a version of NCSM with Woods--Saxon basis  resulted in a value of~2.08(7)~fm, which
agrees with~$r_{p}\approx 2.04$~fm that is the crossing point of JISP16 curves in Fig.~\ref{12Cr}. The crossing of the Daejeon16 curves
suggests~$r_{p}\approx 2.29$~fm that is much closer to the experimental result of~2.32(2)~fm~\cite{AjS90}.


In conclusion, we propose a realistic $NN$ interaction Daejeon16 
based on a SRG-transformed chiral N3LO interaction
that provides a good description of various observables in
light nuclei without $NNN$ forces 
and also generates
rapid convergence in {\em ab initio} calculations. We anticipate that this interaction will
be useful for a wide range of applications to nuclear structure and reactions. 

{\em Acknowledgements}. We thank Gaute Hagen for the SRG-evolved $NN$ interaction and Morten Hjorth-Jensen for use of his codes for SRG-evolution of the $NN$ interaction. This work was supported in part 
by the US Department of Energy under Grants No. DESC0008485 (SciDAC/NUCLEI) and 
No.~DE-FG02-87ER40371, by the US National Science Foundation under Grant No. 1516096,  
by the Russian Foundation for Basic Research Grant No. 15-02-06604-a, and the Ames Laboratory, 
operated by Iowa State University under contract No.~DE-AC02-07CH11358. 
{This work was also partially supported by the Rare Isotope Science Project of Institute for Basic Science funded by Ministry of Science, ICT and Future Planning and National Research Foundation of Korea (2013M7A1A1075764).}
Computational resources were provided by the National Energy Research Supercomputer Center (NERSC), which is supported by the Office of Science of the U.S. Department of Energy under Contract No. DE-AC02-05CH11231.
{A portion of the computational resources were also provided by the Supercomputing Center/Korea Institute of Science and Technology Information including technical support (KSC-2013-C3-052).}

\end{document}